\documentclass{llncs}

\usepackage[T1]{fontenc}
\usepackage[utf8x]{inputenx}

\usepackage{cite}
\usepackage[dvipsnames]{xcolor}
\usepackage{graphicx}
\usepackage{xspace}
\usepackage[hyphens]{url}
\usepackage[breaklinks=true]{hyperref}
\usepackage{booktabs}

\usepackage{array}
\usepackage{enumitem}
\usepackage{upgreek}
\usepackage{multirow}
\usepackage{tabularx}
\usepackage{amssymb}
\usepackage{amsmath}
\usepackage{cleveref}
\usepackage{booktabs}
\usepackage{xifthen}
\usepackage{adjustbox}

\usepackage{floatrow}
\newfloatcommand{capbtabbox}{table}[][\FBwidth]

\usepackage{wrapfig}
\usepackage{tikz}
\usetikzlibrary{calc,automata,positioning,decorations.pathreplacing}
\tikzset{align at top/.style={baseline=(current bounding box.north)}}
\tikzstyle{every node}=[font=\scriptsize]
\tikzstyle{state} = [draw,fill=white,circle,thick,align=center,inner sep=0pt,minimum size=4.5mm]
\tikzstyle{lstate} = [draw,fill=white,rectangle,rounded corners,thick,align=center,inner sep=2pt]
\tikzstyle{dot} = [fill,circle,inner sep=0mm,minimum size=1.25mm,line width=0mm]

\Crefname{figure}{Fig.}{Figs.}
\crefname{figure}{fig.}{figs.}
\Crefname{tabular}{Tab.}{Tabs.}
\crefname{tabular}{tab.}{tabs.}
\Crefname{section}{Sect.}{Sects.}
\crefname{section}{sect.}{sects.}
\Crefname{equation}{Eq.}{Eqs.}
\crefname{equation}{eq.}{eqs.}
\creflabelformat{equation}{#2#1#3}

\setitemize{noitemsep,topsep=0pt,parsep=0pt,partopsep=0pt,leftmargin=12.0pt}
\setenumerate{noitemsep,topsep=0pt,parsep=0pt,partopsep=0pt,leftmargin=12.5pt}
\setdescription{noitemsep,topsep=0pt,parsep=0pt,partopsep=0pt,leftmargin=12.5pt}

\newcommand{\modest}{\textsc{\mbox{Modest}}\xspace}

\newcommand{\tool}[1]{\textsc{#1}}

\newcommand{\eg}{e.g.\ }
\newcommand{\ie}{i.e.\ }
\newcommand{\etal}{et al.\xspace}
\newcommand{\wrt}{w.r.t.\xspace}

\renewcommand{\iff}{\ensuremath{\Leftrightarrow}\xspace}
\newcommand{\set}[1]{\ensuremath{\{\,#1\,\}}}
\newcommand{\tuple}[1]{\ensuremath{\langle #1 \rangle}}

\newcommand{\defeq}{\mathrel{\vbox{\offinterlineskip\ialign{\hfil##\hfil\cr{\tiny \rm def}\cr\noalign{\kern0.30ex}$=$\cr}}}}
\newcommand{\Dist}[1]{\ensuremath{\mathit{Dist}({#1})}\xspace}
\newcommand{\support}[1]{\ensuremath{\mathit{spt}({#1})}\xspace}

\newcommand{\RRplus}{\ensuremath{\mathbb{R}^+}\xspace}
\newcommand{\RRpluszero}{\ensuremath{\mathbb{R}^+_0}\xspace}

\newcommand{\NN}{\ensuremath{\mathbb{N}}\xspace}

\newcommand{\True}[0]{\ensuremath{\mathit{true}}\xspace}
\newcommand{\False}[0]{\ensuremath{\mathit{false}}\xspace}
\newcommand{\xtr}[1]{\xrightarrow{\protect{\raisebox{-1pt}[0pt][0pt]{\ensuremath{\scriptstyle{#1}}}}}}
\newcommand{\blackdiamond}{\raisebox{0.76pt}{\scalebox{0.52}{\rotatebox[origin=c]{45}{\ensuremath{\blacksquare}}}}}

\usepackage[vlined,linesnumbered]{algorithm2e}
\renewcommand{\algorithmcfname}{Alg.}
\SetAlgoCaptionLayout{raggedright}
\SetAlCapFnt{\small}
\SetAlCapNameFnt{\small}
\SetAlCapSkip{\abovecaptionskip\relax}
\SetEndCharOfAlgoLine{}
\SetArgSty{textrm}
\SetKwInOut{Input}{Input}\SetKwInOut{Output}{Output}
\SetCommentSty{textit}
\SetKw{Break}{break}
\SetKwProg{Type}{type}{}{end}
\SetKwProg{Function}{function}{}{end}
\SetKwFor{ForEach}{foreach}{do}{}
\SetKwFor{WhileTrue}{repeat}{}{end}
\SetKw{Continue}{continue}
\SetKwRepeat{DoWhile}{repeat}{while}
\SetKwProg{Fn}{function}{}{}
\crefname{algocf}{alg.}{algs.}
\Crefname{algocf}{Alg.}{Algs.}

\newcommand{\rfunc}{r}
\newcommand{\init}{{\overline{s}}}

\hypersetup{
  pdftitle = {Symblicit Exploration and Elimination for MDPs}
}
\begin{document}

\title{%
Symblicit Exploration and Elimination\\for Probabilistic Model Checking%
\thanks{%
The authors are listed alphabetically.
This work was supported by
NWO VENI grant no.\ 639.021.754.}%
}
\author{
Ernst Moritz Hahn\inst{1}
\and Arnd Hartmanns\inst{2} 
}

\institute{
Queen's University Belfast, Belfast, UK
\and University of Twente, Enschede, The Netherlands}
\date{\today}
\maketitle

\begin{abstract}
Binary decision diagrams can compactly represent vast sets of states, mitigating the state space explosion problem in model checking.
Probabilistic systems, however, require multi-terminal diagrams storing rational numbers.
They are inefficient for models with many distinct probabilities and for iterative numeric algorithms like value iteration.
In this paper, we present a new ``symblicit'' approach to checking Markov chains and related probabilistic models:
We first generate a decision diagram that \emph{symb}olically collects all reachable states and their predecessors.
We then concretise states one-by-one into an exp\emph{licit} partial state space representation.
Whenever all predecessors of a state have been concretised, we \emph{eliminate} it from the explicit state space in a way that preserves all relevant probabilities and rewards.
We thus keep few explicit states in memory at any time.
Experiments show that very large models can be model-checked in this way with very low memory consumption.
\end{abstract}

\section{Introduction}
\label{sec:Introduction}

Many of the complex systems that we are surrounded by, rely on, and use every day are inherently probabilistic:
The Internet is built on randomised algorithms such as the collision avoidance schemes in Ethernet and wireless protocols, with the latter additionally being subject to random message loss.
Hard- and software in cars, trains, and airplanes is designed to be fault-tolerant based on mean-time-to-failure statistics and stochastic wear models.
Machine learning algorithms give recommendations based on estimates of the likelihoods of possible outcomes, which in turn may be learned from randomly sampled data.

Given a formal mathematical model of such a system, \eg in the form of (a high-level description of) a discrete- or continuous-time Markov chain (DTMC or CTMC), probabilistic model checking can automatically compute (an approximation of) the value of a quantity of interest.
Such quantities include the probability to finally reach an unsafe state (a measure of reliability), the steady-state probability to be in a failure state (determining availability), the long-run average reward (measuring \eg throughput or energy consumption), or the accumulated cost up to a certain set of states (where \eg a job is complete).
The standard approach is to proceed in two phases:
First, \emph{explore} the state space, building a representation of the set of reachable states and the transitions connecting them.
Transitions are annotated with rational values for probabilities and rewards, which are usually represented as floating-point numbers.
Second, use an iterative numeric algorithm such as value iteration~\cite{Put94} or one of its sound variants~\cite{HM18,QK18,BKLPW17,HK19} to \emph{compute} the quantity of interest.
These algorithms in fact compute a value for every state that approximates the quantity starting from that state up to a prescribed error~$\epsilon$.
In contrast to classic functional model checking, which admits on-the-fly algorithms for \eg reachability or LTL properties, probabilistic model checking is thus doubly affected by the state space explosion problem:
First, the entire state space must be stored in memory, including many numeric values.
Second, the numeric computation requires multiple vectors of values to be stored, and updates to be performed on them, for all states.

\paragraph{Current approaches} to mitigate the state space explosion problem in probabilistic model checking include the use of partial exploration and learning algorithms, bisimulation minimisation, and compact representations of the state space or value vectors by binary decision diagrams (BDDs).
They exploit different structural properties that only sometimes overlap.
The learning-based approaches~\cite{BCCFKKPU14,ABHK18} for reachability probabilities work well for models where a small initial subset of the state space determines most of the probability mass.
In such cases, which are not abundant among existing case studies~\cite{HHHKKKPQRS19}, they complete in a few seconds while exhaustive approaches run out of time or memory~\cite[Table~1]{BHH19}.
Bisimulation minimisation reduces the state space to a quotient according to a probabilistic bisimulation relation; see \cite[Sect.~5.1]{BHK19} for an overview.
It has been implemented in \tool{Storm}~\cite{DJKV17} and allows certain very large models to be checked efficiently; in general, its impact depends on the amount of bisimilar states in the given system.
Finally, BDDs~\cite{Lee59,Bry18} have a long history of use in (discrete-state) model checking~\cite{CG18} to compactly represent state spaces, in good cases reducing memory usage by orders of magnitude.
They work well when the state space is structured and exhibits symmetry, which is often the case for real-life case studies modelled by humans (as opposed to randomly generated examples).
In probabilistic model checking, however, numeric values from continuous domains are part of state spaces and must be encoded in the decision diagrams.
A binary encoding of their floating-point representation does not usually result in compact diagrams; instead, multi-terminal BDDs (MTBDDs), where each of the (unbounded number of) leaves represents one number, have been applied with some success, notably in the probabilistic model checker \tool{Prism}~\cite{KNP11}.
They however do not provide much compaction for models with many distinct probabilities or reward values due to the large number of leaves.
They also do not work well to represent the large vectors of values used in iterative numeric algorithms such as value iteration, which progress through many very different intermediate values for each state before converging to, but often not reaching, a fixpoint.
For this reason, \tool{Prism} defaults to its \textit{hybrid} engine, which uses MTBDDs for the state space but arrays of double-precision values for iteration.
Its fully symbolic \textit{mtbdd} engine only solves specific large structured models in reasonable runtime.

\paragraph{Our contribution}
is a new approach that combines (MT)BDDs, explicit state representations, and state elimination to tackle the problem of model-checking large probabilistic specifications.
Its novelty lies in
(1)~using BDDs precisely for those tasks that they work well for, and
(2)~using state elimination instead of the standard iterative algorithms in the computation phase.
We work with discrete-state probabilistic systems; in this paper, we use DTMC to explain our approach, but the same techniques apply directly to CTMC, Markov decision processes (MDP)~\cite{Put94}, and Markov automata (MA)~\cite{EHZ10}, too.

Our state space exploration performs a standard explicit-state breadth-first search, but we use a decision diagram instead of the standard hashset to store the set of visited states.
We do not store transitions, thus no continuous numeric values blow the diagram up.
However, we do count the number of predecessors of each state---thus we use an MTBDD.
Since this number is a discrete quantity with low variation in most models, the diagrams usually remain compact.
For the computation phase, we explore the state space again, this time creating a representation that includes transitions, but that is explicit.
During this exploration, we keep track of the number of explored predecessors of each state.
Whenever, for some fully-explored state~$s$, this number reaches the predecessor count given by the MTBDD, we apply \emph{state elimination}:
we remove $s$ from the (yet incomplete) explicit state space representation, and replace all of its incoming and outgoing transitions by direct transitions between the predecessors and successors of~$s$.
By redistributing the original transitions' probabilities and rewards in the right way, the quantity of interest remains unaffected.
This method of computation simultaneously avoids the iterative algorithms' convergence and precision issues~\cite{HM14} and keeps memory usage due to the explicit representation low:
on most models, most states are eliminated soon after they have been fully explored, thus only few need to be kept in memory at any time.
Upon termination, only the initial state and goal state(s) remain, and the value for the quantity of interest can be read off the transitions connecting them.

Two technical insights make our computation phase work well:
First, we use an explicit representation not only to avoid storing (continuous) probabilities and rewards in an MTBDD, but also because state elimination tends to create unstructured intermediate state spaces that would blow up any BDD representation.
Second, the precomputed predecessor count allows us to eliminate a state at precisely the moment after which we will not encounter it again in our search, avoiding costly re-explorations and re-eliminations.

\paragraph{Related work.}
State elimination stems from the classic reduction algorithm to convert a finite automaton into a regular expression~\cite{BM63}.
It was introduced to probabilistic model checking to solve parametric Markov chains~\cite{Daw04} and forms the core of the \tool{Param}~\cite{HHWZ10} and \tool{Prophesy}~\cite{DJJCVBKA15} tools.
For non-parametric models, it enables efficient computation of reward-bounded reachability probabilities~\cite{HH16}.
In this paper, we use it for non-parametric Markov chains and unbounded (infinite-horizon) properties.
In all of these settings, its effectiveness crucially depends on the order in which states are eliminated, which is determined by (configurable) heuristics.
Symblicit techniques have previously been used for long-run average properties~\cite{WBBHCHDT10}, based on bisimulation minimisation, and later expanded in related settings~\cite{BBRB17}.
A different form of elimination on strongly-connected components was used by Gui \etal \cite{GSSLD14} to accelerate the (explicit-state) computation of reachability probabilities via value iteration.

\section{Background}
\label{sec:Background}

\paragraph{Mathematical notions.}
$\RRpluszero$ is the set of all non-negative, and $\RRplus$ the set of all positive, real numbers.
A (discrete) \emph{probability distribution} over $S$ is a function $\mu \in S \to [0, 1]$ with countable \emph{support} $\support{\mu} \defeq \set{ s \in S \mid \mu(s) > 0 }$ and $\sum_{s \in \support{\mu}} \mu(s) = 1$.
$\Dist{S}$ is the set of all probability distributions over $S$. 

\subsection{Discrete-Time Markov Chains}
\label{sec:MarkovChains}

\begin{definition}
\label{def:MarkovChain}
A \emph{discrete-time Markov chain} (DTMC) is a tuple
$M = \tuple{S, s_I, P, R}$
consisting of a
finite set of \emph{states} $S$, an \emph{initial state} $s_I \in S$, a \emph{transition} function $P \colon \mathit{S} \to \Dist{S}$, 
and a \emph{reward} function $R \colon S \to \RRpluszero$.

\end{definition}
We also write a transition as $s \xtr{p} s'$ if $p = P(s)(s') > 0$.
A transition is uniquely identified by the two states it connects.
When in state $s$ of a DTMC, we delay for one time unit before jumping to the next state.
\emph{Continuous-time Markov chains} (CTMC) extend DTMC by additionally assigning a \emph{rate} $Q(s) \in \RRplus$ to every state.
Then the probability to delay for at most $t$ time units is
$1 - \mathrm{e}^{-Q(s) \cdot t}$,
\ie the residence time follows the exponential distribution with rate $Q(s)$.
In both models, the probability to then move to state $s'$ is given by $P(s)$.
When staying for $t$ time units in state $s$, we incur a reward of $R(s) \cdot t$.
To simplify the presentation, we use DTMC throughout this paper, but mention the changes needed in definitions or algorithms to use CTMC, where appropriate.

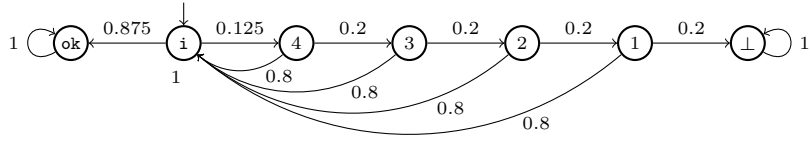
\begin{figure}[t]
\begin{tikzpicture}[on grid,auto]
  \node[state] (i) {\texttt{i}};
  \node[] (ir) [below left=0.45 and 0.1 of i] {$1$};
  \coordinate[above=0.3 of i.north] (start);
  \node[state] (ok) [left=1.5 of i] {\texttt{ok}};
  \node[state] (n4) [right=1.5 of i] {$4$};
  \node[state] (n3) [right=1.5 of n4] {$3$};
  \node[state] (n2) [right=1.5 of n3] {$2$};
  \node[state] (n1) [right=1.5 of n2] {$1$};
  \node[state] (err) [right=1.5 of n1] {$\bot$};
  ;
  \path[->]
    (start) edge node {} (i)
    (i) edge [] node[swap] {$0.875$} (ok)
    (i) edge [] node[] {$0.125$} (n4)
    (n4) edge [] node[] {$0.2$} (n3)
    (n3) edge [] node[] {$0.2$} (n2)
    (n2) edge [] node[] {$0.2$} (n1)
    (n1) edge [] node[] {$0.2$} (err)
    (err) edge [loop,out=-30,in=30,looseness=6] node[swap] {$1$} (err)
    (ok) edge [loop,out=-150,in=150,looseness=6] node[] {$1$} (ok)
    (n4) edge [bend left=40] node[pos=0.25,inner sep=1pt] {$0.8$} (i)
    (n3) edge [bend left=40] node[pos=0.25,inner sep=1pt] {$0.8$} (i)
    (n2) edge [bend left=40] node[pos=0.25,inner sep=1pt] {$0.8$} (i)
    (n1) edge [bend left=40] node[pos=0.25,inner sep=1pt] {$0.8$} (i)
  ;
\end{tikzpicture}
\caption{DTMC $M_z$ for the Zeroconf protocol ($h=32, a=2^8,p=0.2,n=4$)}
\label{fig:ExampleDTMC}
\end{figure}

\begin{example}
\label{ex:MarkovCahin}
As a running example, we use a very abstract model of the Zeroconf protocol~\cite{BSHV03}, shown as DTMC $M_z$ in \Cref{fig:ExampleDTMC} (adapted from~\cite{GHS18}).
We draw transitions as arrows labelled with their probability.
Non-zero rewards are given next to the states.
$M^z$ has 7~states and 12~transitions.
The protocol is used by hosts joining a network to auto-configure a unique IP address.
A new host joining the network of $h = 32$ host starts in state \texttt{i}.
It selects an address uniformly at random from the space of $a = 2^8$ addresses.
The probability that the address is already in use is $\frac{h}{a} = \frac{1}{8}$.
The host checks $n = 4$ times whether its address is already in use.
If it is not, all checks will succeed, modelled by state \texttt{ok}, to which we moved with probability $1 - \frac{h}{a}$.
If it is, then states $n$ down to $1$ model the checks.
Each check can fail to give the correct negative result due to message loss with probability $p = 0.2$.
If all tests do so, then the host incorrectly believes that it has a unique address, in state~$\bot$.
Otherwise, it retries with a newly chosen address from state \texttt{i}.
We incur a reward of $1$ in state \texttt{i}, \ie for every IP address we try.
The size of the DTMC can be blown up arbitrarily via parameter $n$.
\end{example}
In practice, higher-level modelling languages like \modest~\cite{HHHK13} or the \tool{Prism} language~\cite{KNP11} are used to specify larger DTMC.
The semantics of a DTMC is formally captured by its \emph{paths}:

\begin{definition}
Given a DTMC $M$ as above, a \emph{finite path} is a sequence
$\pi_\mathrm{fin} = s_0\, t_0\, s_1\, t_1 \dots s_n$
of states $s_i \in S$ and delays $t_i \in \RRplus$ where $P(s_i)(s_{i + 1}) > 0$ and $t_i = 1$ for all $i \in \set{ 0, \dots, n - 1 }$.
Let $|\pi_\mathrm{fin}| \defeq n$,
$\mathrm{last}({\pi_\mathrm{fin}}) \defeq s_n$,
$\mathrm{dur}(\pi_\mathrm{fin}) \defeq \sum_{i=0}^{n-1} t_i$,
and $\mathrm{rew}({\pi_\mathrm{fin}}) \defeq \sum_{i=0}^{n-1} t_i \cdot R(s_i)$.
$\Pi_\mathit{fin}$ is the set of all finite paths starting in $s_I$.
A \emph{path} is an analogous infinite sequence $\pi$, and $\Pi$ are all paths starting in $s_I$.
We define $s \in \pi \iff \exists\, i \colon s = s_i$.
Let $\pi_{\to m}$ for $m \in \NN$ be the prefix of $\pi$ of length $m$, \ie $|\pi_{\to m}| = m$, and let $\pi_{\to G}$ be the shortest prefix of $\pi$ that contains a state in $G \subseteq S$, or $\bot$ if $\pi$ contains no such state.
Let $\mathrm{rew}(\bot) \defeq \infty$.
\end{definition}
In CTMC, the $t_i$ can be arbitrary numbers in $\RRpluszero$.
For $M$ as above, following the rules described below \Cref{def:MarkovChain} and the standard cylinder set construction~\cite{BK08}, we obtain a probability measure $\mathbb{P}_M$ on measurable sets of paths starting in $s_I$.

\begin{definition}
Given a set of goal states $G \subseteq S$, the \emph{reachability probability} \wrt $g$ is $\mathbb{P}(\diamond\: G) \defeq \mathbb{P}_M( \pi \in \Pi \mid \exists\, g \in G \colon g \in \pi)$.
Let $r_G \colon \Pi \to \RRpluszero$ be the random variable defined by $r_G(\pi) = \mathrm{rew}(\pi_{\to G})$.
Then the \emph{expected reward} to reach $G$ is the expected value of $r_G$ under $\mathbb{P}_M$, written as $\mathbb{E}(\blackdiamond\: G)$.
Let $r_\mathit{lra} \colon \Pi \to \RRpluszero$ be defined by $r_\mathit{lra}(\pi) = \lim_{i \to \infty} \mathrm{rew}(\pi_{\to i})/\mathrm{dur}(\pi_{\to i})$.
Then the \emph{long-run average reward} is the expected value of $r_\mathit{lra}$ under $\mathbb{P}_M$, written as $\mathbb{L}$.
\end{definition}
The steady-state probability $\mathbb{S}(S')$ of residing in a state in $S' \subseteq S$ is a special case of the long-run average reward where $R(s) = 1$ if $s \in S'$ and $0$ otherwise.
Whenever we consider a DTMC with a set of goal states $G$, we assume that they have been made absorbing, \ie that for all $g \in G$ we have $P(g)(g) = 1$.
Given a CTMC, reachability probabilities and expected rewards can be computed on its \emph{embedded DTMC}, obtained by dividing all rewards by $Q(s)$; only for long-run averages do we need a dedicated treatment of the rates resp.\ residence times.

\begin{example}
For our Zeroconf example DTMC $M_z$ from \Cref{fig:ExampleDTMC}, we may want to compute the probability to eventually pick a unique address $\mathbb{P}(\diamond\: \set{\texttt{ok}})$, which will be just below $1$, and the expected number of addresses that we ever try $\mathbb{E}(\blackdiamond\: \set{ \texttt{ok}, \bot })$.
Note that $\mathbb{E}(\blackdiamond\: \set{ \texttt{ok} })$ is $\infty$ by definition since the set of paths that never reach state \texttt{ok} has positive probability.
\end{example}

\subsection{Binary Decision Diagrams}
\label{sec:BDDs}

\emph{Binary decision diagrams} (BDDs)~\cite{Lee59,Bry18} represent Boolean functions as rooted directed acyclic graphs.
They have two leaf nodes, \True and \False.
Every inner node is associated to one input bit, and has two children: the high (solid line) and low (dotted line) child.
On a path from the root to a leaf, every bit must occur at most one.
Such a path corresponds to the inputs in which bit $b_i$ is assigned to \True (\False) if we go from a node for $b_i$ to its high (low) child.
We typically order the bits on all paths, merge isomorphic subgraphs, and remove redundant nodes.
Then a BDD can represent many functions with few nodes.

In model checking, BDDs are used to represent sets of states (by assigning \True to the binary encoding of a state iff it is in the set) as well as the transition relation (by assigning \True to the binary encoding of a pair of states if they are connected by a transition).
In probabilistic model checking, however, we need to encode functions that map to rational numbers to encode transition probabilities, rewards, and the value vectors in value iteration.
Most tools represent them as 64-bit floating-point values, but the corresponding binary representation does not typically allow good compression with BDDs.
Symbolic probabilistic model checkers such as \tool{Prism}~\cite{KNP11} this use \emph{multi-terminal} BDDs (MTBDDs) with one leaf node per number.
Since a finite model only contains finitely many probabilities, or values for states, this approach is effective, but often not efficient:
for example, when performing value iteration on our example DTMC $M_z$ for $\mathbb{P}(\diamond\: \set{\texttt{ok}})$, we have to encode the following function after 5 iterations:\\[2pt]
\centerline{%
$\set{ \texttt{i} \mapsto 0.99225, 4 \mapsto 0.9716, 3 \mapsto 0.966, 2 \mapsto 0.938, 1 \mapsto 0.784, \texttt{ok} \mapsto 1, \bot \mapsto 0 }$%
}\\[2pt]
Observe that every state has a distinct value, thus the MTBDD offers no compression.
In practice, they only work well for very specific models with few distinct transition probabilities and rewards, and where the iterative numeric algorithms assign the same (intermediate) values to many states.

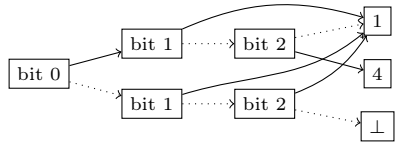
\begin{figure}[t]
\begin{tikzpicture}[on grid,auto]
  \node[draw=black] (b0) {bit 0};
  \node[draw=black] (b1_0) [above right=0.4 and 1.5 of b0] {bit 1};
  \node[draw=black] (b1_1) [below right=0.4 and 1.5 of b0] {bit 1};
  \node[draw=black] (b2_1) [below right=0.0 and 1.5 of b1_0] {bit 2};
  \node[draw=black] (b2_3) [right=1.5 of b1_1] {bit 2};
  \node[draw=black] (r1) [above right=0.3 and 1.5 of b2_1] {$1$};
  \node[draw=black] (r4) [below right=0.4 and 1.5 of b2_1] {$4$};
  \node[draw=black] (rbot) [below right=0.3 and 1.5 of b2_3] {$\bot$};
  ;
  \path[->]
    (b0) edge [] (b1_0)
    (b0) edge [dotted] (b1_1)
    (b1_0) edge [bend left=20] (r1)
    (b1_0) edge [dotted] (b2_1)
    (b1_1) edge [out=17,in=-140] (r1)
    (b1_1) edge [dotted] (b2_3)
    (b2_3) edge [bend right=17] (r1)
    (b2_3) edge [dotted] (rbot)
    (b2_1) edge [dotted] (r1)
    (b2_1) edge [] (r4)
  ;
\end{tikzpicture}
\caption{MTBDD counting the numbers of predecessors for all states of $M_z$}
\label{fig:ExampleMTBDD}
\end{figure}

\begin{example}
\label{ex:MTBDD}
\Cref{fig:ExampleMTBDD} shows an MTBDD mapping every state of $M_z$ to its number of predecessor states.
We have 7 states, thus use 3 bits for their encoding.
States $1$ through $4$ are encoded as that number, \texttt{i} is $5$ ($101_\mathrm{2}$), \texttt{ok} is $6$ (($110_\mathrm{2}$), and $\bot$ is $7$ ($111_\mathrm{2}$).
There is no (reachable) state encoded as $0$, thus we map $0$ to the extra $\bot$ leaf node---in this way, such an MTBDD can indicate that certain states are unreachable, or have not been explored yet.
Observe that the MTBDD representation achieves some compression by excluding two redundant nodes for bit~2.
If we scale the model up by increasing $n$, the compression would increase.
\end{example}

\subsection{State Elimination}
\label{sec:StateElimination}

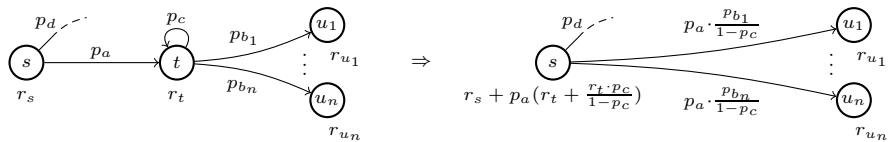
\begin{figure}[t]
\begin{tikzpicture}[on grid,auto]
  \node[state] (s) {$s$};
  \coordinate[above right=0.33 and 0.33 of s] (off1);
  \coordinate[above right=0.6 and 0.8 of s] (off2);
  \node[state] (t) [right=2 of s] {$t$};
  \node[state] (u1) [above right=0.5 and 2 of t] {$u_1$};
  \node[] (udots) [right=1.7 of t] {\rotatebox{90}{$\dots$}};
  \node[state] (un) [below right=0.5 and 2 of t] {$u_n$};
  \node[] (to) [right=5.25 of s] {$\Rightarrow$};
  \node[] (sr) [below=0.45 of s] {$r_s$};
  \node[] (tr) [below=0.45 of t] {$r_t$};
  \node[] (u1r) [below right=0.45 and 0.25 of u1] {$r_{u_1}$};
  \node[] (unr) [below right=0.45 and 0.25 of un] {$r_{u_n}$};
  \node[state] (sp) [right=7 of s] {$s$};
  \coordinate[above right=0.33 and 0.33 of sp] (off1p);
  \coordinate[above right=0.6 and 0.8 of sp] (off2p);
  \node[state] (u1p) [above right=0.5 and 4 of sp] {$u_1$};
  \node[] (udotsp) [right=3.7 of sp] {\rotatebox{90}{$\dots$}};
  \node[state] (unp) [below right=0.5 and 4 of sp] {$u_n$};
  \node[] (spr) [below=0.45 of sp] {$r_s + p_a ( r_t + \frac{r_t \cdot p_c}{1-p_c} )$};
  \node[] (u1pr) [below right=0.45 and 0.25 of u1p] {$r_{u_1}$};
  \node[] (unpr) [below right=0.45 and 0.25 of unp] {$r_{u_n}$};
  ;
  \path[-]
    (s) edge[] node {} (off1)
    (off1) edge[densely dashed,bend left=17.5] node[pos=0.25,inner sep=0.5pt] {$p_d$} (off2)
    (sp) edge[] node {} (off1p)
    (off1p) edge[densely dashed,bend left=17.5] node[pos=0.25,inner sep=0.5pt] {$p_d$} (off2p)
  ;
  \path[->]
    (s) edge[] node[pos=0.5,inner sep=1.5] {$p_a$} (t)
    (t) edge[loop,out=60,in=120,looseness=4] node[swap,pos=0.5,inner sep=1.5pt] {$p_c$} (t)
    (t) edge[bend right=10] node[pos=0.6,inner sep=1.5pt] {$p_{b_1}$} (u1)
    (t) edge[bend left=10] node[swap,pos=0.6,inner sep=1.5pt] {$p_{b_n}$} (un)
    (sp) edge[bend right=5] node[pos=0.75,inner sep=1.5pt] {$p_a{\cdot}\frac{p_{b_1}}{1-p_c}$} (u1p)
    (sp) edge[bend left=5] node[swap,pos=0.75,inner sep=1.5pt] {$p_a{\cdot}\frac{p_{b_n}}{1-p_c}$} (unp)
  ;
\end{tikzpicture}\vspace{7.85pt}
\caption{DTMC state elimination}
\label{fig:Elimination}
\end{figure}

State elimination is a process by which a state of a DTMC is removed, \ie transitions are modified such that it is no longer reachable from the initial state and it is removed from the state set $S$, in a way that preserves the values of all properties of interest.
We show the schematic of state elimination in \Cref{fig:Elimination}:
we eliminate a state $t$ by redistributing the probability to enter a self-loop onto its other outgoing transitions, then combine its incoming and outgoing transitions.
It is easy to see that this preserves the probabilities of all measurable sets of paths that pass through $t$ when projecting $t$ out from every path.
In particular, the paths that forever take the self-loop have probability mass zero, which is why we could eliminate the loop.
For rewards, the transformation only preserves the \emph{expected} reward values of sets of paths:
\begin{enumerate}
\item 
In $t$, the expected number of times we take the self-loop is $\frac{p_c}{1 - p_c}$, thus the expected reward from passing through $t$ is $\frac{r_t \cdot p_c}{1 - p_c}$ (for the loop) plus $r_t$ (for taking one of the other outgoing transitions.
\item
Out of $s$, the probability to enter $t$ next is $p_a$, thus we multiply the expected reward of passing through $t$ by $p_a$ and add this to the reward of $s$.
\end{enumerate}
\Cref{alg:Eliminate} shows the pseudocode to perform state elimination on a DTMC stored in explicit data structures (\ie hash sets for states, lists of transitions, etc.).

\begin{algorithm}[t]
\Function(\tcp*[f]{all explicit}){\texttt{Eliminate}$(\tuple{S, s_I, P, R}$, $s \in S$, $S_\mathit{keep} \subseteq S)$}{
  \If(\tcp*[f]{if $s$ has a self-loop and other transitions:}){$s \in \support{P(s)} \wedge P(s) \,{<}\, 1$}{
    \ForEach(\tcp*[f]{redistribute onto other transitions}){$s' \in \support{P(s)} \setminus \set{ s }$}{
      $P(s)(s') := P(s)(s') / (1 - P(s)(s))$\tcp*{the self-loop's probability}
    }
    $R(s) := R(s) + R(s) \cdot {P(s)(s)}/({1 - P(s)(s)})$\tcp*{add the expected reward}
    $P(s)(s) := 0$\tcp*{remove the self-loop}
  }
  \ForEach(\tcp*[f]{for every predecessor $s_\mathit{pre}$:}){$s_\mathit{pre} \in \set{ s'' \mid s \in \support{P(s'')} } \setminus \set{ s }$}{
    $p := P(s_\mathit{pre})(s)$, $P(s_\mathit{pre})(s) := 0$\tcp*{remove the transition from $s_\mathit{pre}$ to $s$}
    \ForEach(\tcp*[f]{then merge the transitions of $s$}){$s' \in \support{P(s)}$}{
      $P(s_\mathit{pre})(s') := P(s_\mathit{pre})(s') + p \cdot P(s)(s')$\tcp*{into the transitions of~$s_\mathit{pre}$}
    }
    $R(s_\mathit{pre}) := R(s_\mathit{pre}) + p \cdot R(s)$\tcp*{merge the reward of $s$ into that of $s_\mathit{pre}$}
  }
    \If(\tcp*[f]{if $s$ is not needed: remove}){$s \notin S_\mathit{keep} \wedge P(s)(s) = 0$}{
      $P := P \setminus \set{ s \mapsto P(s) }$, $R := R \setminus \set{ s \mapsto R(s) }$\tcp*{its transitions, reward,}
      $S := S \setminus \set{ s }$\tcp*{and the state itself}
    }
}
\caption{State elimination for probabilities and expected rewards}
\label{alg:Eliminate}
\end{algorithm}

\section{Symblicit Exploration and Elimination}
\label{sec:StateElimination}

As we explained in \Cref{sec:Introduction} and illustrated in \Cref{sec:BDDs}, many probabilistic models do not give rise to a compact BDD-based representation if the numeric values---probabilities, rewards, rates for CTMC---are included.
Furthermore, the standard iterative numeric algorithms like value iteration usually produce data that is hardly BDD-compressible.
In this section, we present a combined symbolic-explicit approach that uses MTBDDs in a way that usually avoids such problems, and that uses state elimination to calculate $\mathbb{P}$, $\mathbb{E}$, and $\mathbb{L}$ values without having to keep (values for all states of) the entire state space in memory.

\begin{algorithm}[t]
\Function(\tcp*[f]{explicit $s_I$, executable \texttt{P}, \texttt{R}, \texttt{G}}){\texttt{ExploreEliminate}$(s_I$, \texttt{P}, \texttt{R}, \texttt{G}$)$}{
  $\hat{\mathit{pre}} := \texttt{Explore}(s_I$, \texttt{P}$)$\label{alg:ExploreEliminate:Explore}\tcp*{get predecessor count MTBDD for all states}
  $\mathit{done} := \varnothing$, $\mathit{agenda} := \set{ s_I }$\label{alg:ExploreEliminate:Rest}\tcp*{$\mathit{done}$ stored as hash set, $\mathit{agenda}$ as queue}
  $S := \set{ s_I }$, $P := \varnothing$, $R := \varnothing$, $\mathit{pre}' := \set{ s_I \mapsto 0 }$\tcp*{explicit sets and functions}
  \While(\label{alg:ExploreEliminate:While}){$\mathit{agenda} \neq \varnothing$}{
    $s := \text{next element of } \mathit{agenda}$, $\mathit{agenda} := \mathit{agenda} \setminus \set{ s }$\;
    \ForEach(\tcp*[f]{explore $s$}){$s' \in \support{\texttt{P}(s)}$}{
      $P(s)(s') := \texttt{P}(s)(s')$, $R(s) := \texttt{R}(s)$\;
      \If{$s' \notin S$}{
        $S := S \cup \set{ s }$, $\mathit{agenda} := \mathit{agenda} \cup \set{ s }$\;
        $\mathit{pre}' := \mathit{pre}' \cup \set{ s' \mapsto 0 }$
      }
      \lIf{$s' \neq s$}{$\mathit{pre}'(s') := \mathit{pre}'(s') + 1$}
    }
    $\mathit{done} := \mathit{done} \cup \set{ s }$\label{alg:ExploreEliminate:FullyExplored}\tcp*{$s$ is now fully explored}
    $E := \set{ s_e \mid s_e \in \{ s \} \,{\cup}\, \support{P(s)} \,{\cap}\, \mathit{done} }$\tcp*{collect just modified states}
    $E := \set{ s_e \mid s_e \in E \wedge \hat{\mathit{pre}}(s_e) \,{=}\, \mathit{pre}'(s) }$\tcp*{with all predecessors explored}
    \ForEach(\tcp*[f]{and eliminate them}){$s_\mathit{elim} \in E$}{
      \texttt{Eliminate}$(\tuple{S, s_I, P, R}, s_\mathit{elim}, \set{ s_I })$\label{alg:ExploreEliminate:Eliminate}\;
      \If(\tcp*[f]{cleanup}){$s_\mathit{elim} \notin S$}{%
        $\mathit{pre}' := \mathit{pre}' \setminus \set{ s_\mathit{elim} \mapsto \mathit{pre}'(s_\mathit{elim}) }$,
        $\mathit{done} := \mathit{done} \setminus \set{ s_\mathit{elim} }$
      }
    }
  }
  \lIf{we compute a reachability probability}{\Return{%
    $\sum_{g \in \texttt{G}} P(s_I)(g)$\label{alg:ExploreEliminate:PReturn}
  }}
  \ElseIf{we compute an expected reward}{
    \lIf(\tcp*[f]{we have $\mathbb{P}(\diamond\: \texttt{\emph{G}}) < 1$}){$\support{P(s_I)} \setminus \texttt{G} \neq \varnothing$}{\Return{$\infty$\label{alg:ExploreEliminate:EReturnInfty}}}
    \lElse{\Return{$R(s_i) + \sum_{g \in \texttt{G}} P(s_I)(g) \cdot R(g)$\label{alg:ExploreEliminate:EReturnFinite}}}
  }
}
\caption{Symblicit exploration-elimination for probabilities and expected rewards}
\label{alg:ExploreEliminate}
\end{algorithm}

The pseudocode of our approach is shown as function \texttt{ExploreEliminate} in \Cref{alg:ExploreEliminate}.
It uses functions \texttt{Explore} of \Cref{alg:Explore} and \texttt{Eliminate} of \Cref{alg:Eliminate}.
We typeset values that represent executable code in \texttt{monospace} font:
compact specifications in high-level modelling languages are typically compiled to or interpreted as functions that, given an explicit (bit string) representation of a state, enumerate its transitions (\texttt{P}), compute its reward (\texttt{R}), and return \True iff it is a goal state (\texttt{G}).
We mark variables storing symbolic data (\ie BDDs or MTBDDs) with a $\hat{\mathit{hat}}$.
All other values typeset in $\mathit{italics}$ use explicit data structures such as bit strings for states, hash sets or queues of such bit strings, lists of transitions, etc.

\begin{algorithm}[t]
\Function(\tcp*[f]{explicit $s_I$, executable \texttt{P}}){\texttt{Explore}$(s_I$, \texttt{P}$)$}{
  $\hat{\mathit{seen}} := \set{ s_I }$, $\mathit{agenda} := \set{ s_I }$\tcp*{$\mathit{seen}$ stored as BDD, $\mathit{agenda}$ as queue}
  $\hat{\mathit{pre}} := \set{ s_I \mapsto 0 }$\tcp*{predecessor count, stored as MTBDD}
  \While{$\mathit{agenda} \neq \varnothing$}{
    $s := \text{next element of } \mathit{agenda}$, $\mathit{agenda} := \mathit{agenda} \setminus \set{ s }$\;
    \ForEach{$s' \in \support{\texttt{P}(s)} \setminus \set{ s }$}{
      \If{$s' \notin \hat{\mathit{seen}}$}{
        $\hat{\mathit{seen}} := \hat{\mathit{seen}} \cup \set{ s }$, $\mathit{agenda} := \mathit{agenda} \cup \set{ s }$\;
        $\hat{\mathit{pre}} := \hat{\mathit{pre}} \cup \set{ s' \mapsto 0 }$
      }
      $\hat{\mathit{pre}}(s') := \hat{\mathit{pre}}(s') + 1$\tcp*{$s$ is a previously-unseen predecessor of $s'$}
    }
  }
  \Return{$\hat{\mathit{pre}}$}
}
\caption{Symbolic exploration via  breadth-first search with predecessor counting}
\label{alg:Explore}
\end{algorithm}

Our first step, in line~\ref{alg:ExploreEliminate:Explore}, is to symbolically explore the set of reachable states by calling function \texttt{Explore}.
This function performs a standard breadth-first search, using a BDD for the set of visited states, and additionally constructs an MTBDD that counts the number of predecessors of each state like the one shown in \Cref{fig:ExampleMTBDD} for $M_z$.
In our implementation, $\mathit{seen}$ and $\mathit{pre}$ are actually managed in a single MTBDD as explained in \Cref{ex:MTBDD}.

We then, starting from line~\ref{alg:ExploreEliminate:Rest}, perform another exploration of the state space.
This time, however, we use explicit data structures, and we track the number of \emph{fully explored} predecessors for every state in hash table $\mathit{pre}'$.
A state is fully explored if its reward, all of its transitions, and all successor states, have been added to the explicit representations for $S$, $R$, and $P$.
We track the set of fully explored states in hash set $\mathit{done}$.
Whenever we are done visiting a state $s$ in this second exploration (\ie in line~\ref{alg:ExploreEliminate:FullyExplored} and below), it has just become fully explored, and the fully-explored-predecessor count of its successors has changed.
We then check which of these changed states fulfils the criteria for being eliminated:
It must be fully explored (which only $s$ is for certain), and all of its predecessors must be fully explored (which we determine by comparing $\mathit{pre}'$ and $\hat{\mathit{pre}}$).
We call \texttt{Eliminate} on these states in line~\ref{alg:ExploreEliminate:Eliminate}.
In this way, if we indeed manage to eliminate most states soon after they have been explored, the explicit data structures---$S$, $P$, $R$, $\mathit{pre}'$, $\mathit{done}$, etc.---only track few states at any time and thus consume little memory.
The predecessor count in $\hat{\mathit{pre}}$ is crucial for being able to perform efficient elimination; without it, we would have to apply heuristics that could lead to states being eliminated that would later be explored as successors of other states again, leading to costly re-exploration and re-eliminations.

In \texttt{Eliminate}, if a state is part of the set $S_\mathit{keep}$, we still modify and ``redirect'' the transitions of its predecessors to go around this state, but we do not remove it from the state space.
We use this to avoid eliminating the initial state $s_I$.
We also do not eliminate states whose only transition is a self-loop: they do not have successors to which transitions could be redirected.
Elimination will thus eventually reduce each bottom strongly connected component (BSCC) of the DTMC to one such self-loop state.
Since we assume all goal states to only have a self-loop, each of them is a BSCC.
Once the outer loop of line~\ref{alg:ExploreEliminate:While} in \texttt{ExploreEliminate} finishes, \texttt{Eliminate} has been called for all states.
Every surviving state at this point is thus the result of eliminating a number of transient states plus a non-goal BSCC or a goal state, and has become a direct successor of the initial state.
We can then directly read the value of $\mathbb{P}(\diamond\: \texttt{G})$ from the transitions to the goal states (line~\ref{alg:ExploreEliminate:PReturn}).
Similarly, the value of $\mathbb{E}(\blackdiamond\: \texttt{G})$ can be derived directly from the remaining rewards, if it is not $\infty$ by definition (lines \ref{alg:ExploreEliminate:EReturnInfty}-\ref{alg:ExploreEliminate:EReturnFinite}).

\begin{figure}[t]
\usetikzlibrary{backgrounds}
\begin{adjustbox}{valign=t,minipage={.525\textwidth}}
\begin{tikzpicture}[on grid,auto]
  \node[state] (i) {\texttt{i}};
  \node[] (ir) [below left=0.45 and 0.1 of i] {$1$};
  \coordinate[above=0.3 of i.north] (start);
  \node[state,dashed] (ok) [left=1.5 of i] {\texttt{ok}};
  \node[] (label) [above left=0.45 and 0.75 of ok] {$(1)$};
  \node[state,dashed] (n4) [right=1.5 of i] {$4$};
  ;
  \path[->]
    (start) edge node {} (i)
    (i) edge [] node[swap] {$0.875$} (ok)
    (i) edge [] node[] {$0.125$} (n4)
  ;
\end{tikzpicture}
\end{adjustbox}%
\begin{adjustbox}{valign=t,minipage={.475\textwidth}}
\begin{tikzpicture}[on grid,auto]
  \node[state] (i) {\texttt{i}};
  \node[] (ir) [below left=0.45 and 0.1 of i] {$1$};
  \coordinate[above=0.3 of i.north] (start);
  \node[state] (ok) [left=1.5 of i] {\texttt{ok}};
  \node[] (label) [above left=0.45 and 0.75 of ok] {$(2)$};
  \node[state,dashed] (n4) [right=1.5 of i] {$4$};
  ;
  \path[->]
    (start) edge node {} (i)
    (i) edge [] node[swap] {$0.875$} (ok)
    (i) edge [] node[] {$0.125$} (n4)
    (ok) edge [loop,out=-150,in=150,looseness=6] node[] {$1$} (ok)
  ;
\end{tikzpicture}
\end{adjustbox}
\begin{adjustbox}{valign=t,minipage={.525\textwidth}}
\begin{tikzpicture}[on grid,auto]
  \node[state] (i) {\texttt{i}};
  \node[] (ir) [below left=0.45 and 0.1 of i] {$1$};
  \coordinate[above=0.3 of i.north] (start);
  \node[state] (ok) [left=1.5 of i] {\texttt{ok}};
  \node[] (label) [above left=0.45 and 0.75 of ok] {$(3)$};
  \node[state] (n4) [right=1.5 of i] {$4$};
  \node[state,dashed] (n3) [right=1.5 of n4] {$3$};
  ;
  \path[->]
    (start) edge node {} (i)
    (i) edge [] node[swap] {$0.875$} (ok)
    (i) edge [] node[] {$0.125$} (n4)
    (n4) edge [] node[] {$0.2$} (n3)
    (ok) edge [loop,out=-150,in=150,looseness=6] node[] {$1$} (ok)
    (n4) edge [bend left=40] node[pos=0.25,inner sep=1pt] {$0.8$} (i)
  ;
\end{tikzpicture}
\end{adjustbox}%
\begin{adjustbox}{valign=t,minipage={.475\textwidth}}
\begin{tikzpicture}[on grid,auto]
  \node[state] (i) {\texttt{i}};
  \node[] (ir) [below left=0.45 and 0.1 of i] {$1$};
  \coordinate[above=0.3 of i.north] (start);
  \node[state] (ok) [left=1.5 of i] {\texttt{ok}};
  \node[] (label) [above left=0.45 and 0.75 of ok] {$(4)$};
  \node[state,dashed] (n3) [right=1.5 of i] {$3$};
  ;
  \path[->]
    (start) edge node {} (i)
    (i) edge [] node[swap] {$0.875$} (ok)
    (i) edge [] node[] {$0.025$} (n3)
    (ok) edge [loop,out=-150,in=150,looseness=6] node[] {$1$} (ok)
    (i) edge [loop,out=-10,in=-70,looseness=6] node[pos=0.5,inner sep=1pt] {$0.1$} (i)
  ;
\end{tikzpicture}
\end{adjustbox}
\begin{adjustbox}{valign=t,minipage={.525\textwidth}}
\begin{tikzpicture}[on grid,auto]
  \node[state] (i) {\texttt{i}};
  \node[] (ir) [below left=0.25 and 0.35 of i] {$1$};
  \coordinate[above=0.3 of i.north] (start);
  \node[state] (ok) [left=1.5 of i] {\texttt{ok}};
  \node[] (label) [above left=0.45 and 0.75 of ok] {$(5)$};
  \node[state] (n4) [right=1.5 of i] {$3$};
  \node[state,dashed] (n3) [right=1.5 of n4] {$2$};
  ;
  \path[->]
    (start) edge node {} (i)
    (i) edge [] node[swap] {$0.875$} (ok)
    (i) edge [] node[] {$0.025$} (n4)
    (n4) edge [] node[] {$0.2$} (n3)
    (ok) edge [loop,out=-150,in=150,looseness=6] node[] {$1$} (ok)
    (n4) edge [bend left=40] node[pos=0.25,inner sep=1pt] {$0.8$} (i)
    (i) edge [loop,out=-60,in=-120,looseness=6] node[pos=0.5] {$0.1$} (i)
  ;
\end{tikzpicture}
\end{adjustbox}%
\begin{adjustbox}{valign=t,minipage={.475\textwidth}}
\begin{tikzpicture}[on grid,auto]
  \node[state] (i) {\texttt{i}};
  \node[] (ir) [below left=0.45 and 0.1 of i] {$1$};
  \coordinate[above=0.3 of i.north] (start);
  \node[state] (ok) [left=1.5 of i] {\texttt{ok}};
  \node[] (label) [above left=0.45 and 0.75 of ok] {$(6)$};
  \node[state,dashed] (n3) [right=1.5 of i] {$2$};
  ;
  \path[->]
    (start) edge node {} (i)
    (i) edge [] node[swap] {$0.875$} (ok)
    (i) edge [] node[] {$0.005$} (n3)
    (ok) edge [loop,out=-150,in=150,looseness=6] node[] {$1$} (ok)
    (i) edge [loop,out=-10,in=-70,looseness=6] node[pos=0.5,inner sep=1pt] {$0.12$} (i)
  ;
\end{tikzpicture}
\end{adjustbox}\\[-0pt]
$\dots$\\[6pt]
\begin{adjustbox}{valign=t,minipage={.525\textwidth}}
\begin{tikzpicture}[on grid,auto]
  \node[state] (i) {\texttt{i}};
  \node[] (ir) [below left=0.25 and 0.35 of i] {$1$};
  \coordinate[above=0.3 of i.north] (start);
  \node[state] (ok) [left=1.5 of i] {\texttt{ok}};
  \node[] (label) [above left=0.45 and 0.75 of ok] {$(9)$};
  \node[state] (n4) [right=1.5 of i] {$1$};
  \node[state,dashed] (n3) [right=1.5 of n4] {$\bot$};
  ;
  \path[->]
    (start) edge node {} (i)
    (i) edge [] node[swap] {$0.875$} (ok)
    (i) edge [] node[] {$0.001$} (n4)
    (n4) edge [] node[] {$0.2$} (n3)
    (ok) edge [loop,out=-150,in=150,looseness=6] node[] {$1$} (ok)
    (n4) edge [bend left=40] node[pos=0.25,inner sep=1pt] {$0.8$} (i)
    (i) edge [loop,out=-60,in=-120,looseness=6] node[pos=0.5] {$0.124$} (i)
  ;
\end{tikzpicture}
\end{adjustbox}%
\begin{adjustbox}{valign=t,minipage={.475\textwidth}}
\begin{tikzpicture}[on grid,auto]
  \node[state] (i) {\texttt{i}};
  \node[] (ir) [below left=0.25 and 0.35 of i] {$1$};
  \coordinate[above=0.3 of i.north] (start);
  \node[state] (ok) [left=1.5 of i] {\texttt{ok}};
  \node[] (label) [above left=0.45 and 0.75 of ok] {$(10)$};
  \node[state] (n4) [right=1.5 of i] {$1$};
  \node[] (n4r) [above=0.45 of n4] {$\frac{200}{219}$};
  \node[state,dashed] (n3) [right=1.5 of n4] {$\bot$};
  ;
  \path[->]
    (start) edge node {} (i)
    (i) edge [] node[swap] {$\frac{875}{876}$} (ok)
    (i) edge [] node[] {$\frac{1}{876}$} (n4)
    (n4) edge [] node[] {$0.2$} (n3)
    (ok) edge [loop,out=-150,in=150,looseness=6] node[] {$1$} (ok)
    (n4) edge [bend left=30] node[pos=0.4,inner sep=2pt] {$\frac{175}{219}$} (ok)
    (n4) edge [loop,out=-60,in=-120,looseness=6] node[pos=0.5] {$\frac{1}{1095}$} (n4)
  ;
\end{tikzpicture}
\end{adjustbox}
\begin{adjustbox}{valign=t,minipage={.525\textwidth}}
\begin{tikzpicture}[on grid,auto]
  \node[state] (i) {\texttt{i}};
  \node[] (ir) [below=0.45 of i] {$1{+}\frac{119918}{119793}$};
  \coordinate[above=0.3 of i.north] (start);
  \node[state] (ok) [left=1.5 of i] {\texttt{ok}};
  \node[] (label) [above left=0.45 and 0.75 of ok] {$(11)$};
  \node[state,dashed] (n4) [right=1.5 of i] {$\bot$};
  ;
  \path[->]
    (start) edge node {} (i)
    (i) edge [] node[swap] {$\frac{4375}{4376}$} (ok)
    (i) edge [] node[] {$\frac{1}{4376}$} (n4)
    (ok) edge [loop,out=-150,in=150,looseness=6] node[] {$1$} (ok)
  ;
\end{tikzpicture}
\end{adjustbox}%
\begin{adjustbox}{valign=t,minipage={.475\textwidth}}
\begin{tikzpicture}[on grid,auto]
  \node[state] (i) {\texttt{i}};
  \node[] (ir) [below=0.45 of i] {$1{+}\frac{119918}{119793}$};
  \coordinate[above=0.3 of i.north] (start);
  \node[state] (ok) [left=1.5 of i] {\texttt{ok}};
  \node[] (label) [above left=0.45 and 0.75 of ok] {$(12)$};
  \node[state] (n4) [right=1.5 of i] {$\bot$};
  ;
  \path[->]
    (start) edge node {} (i)
    (i) edge [] node[swap] {$\frac{4375}{4376}$} (ok)
    (i) edge [] node[] {$\frac{1}{4376}$} (n4)
    (ok) edge [loop,out=-150,in=150,looseness=6] node[] {$1$} (ok)
    (n4) edge [loop,out=-30,in=30,looseness=6] node[swap] {$1$} (n4)
  ;
\end{tikzpicture}
\end{adjustbox}
\caption{Example for exploration with interleaved elimination on $M_z$}
\label{fig:ExampleAlg}
\end{figure}

\begin{example}
For our example DTMC $M_z$ of \Cref{fig:ExampleDTMC}, we have already shown the predecessor count MTBDD computed by \texttt{Explore} in \Cref{fig:ExampleMTBDD}.
Let us now step through the rest of \texttt{ExploreEliminate} on this model.
The partial state spaces that we consider in each step are shown in \Cref{fig:ExampleAlg}.
Fully explored states are drawn with solid outlines, all other states (\ie those in $S$ but not in $\mathit{done}$) with dashed outlines.
In step~(1), we have  just fully explored state \texttt{i}, \ie we just executed line~\ref{alg:ExploreEliminate:FullyExplored} in the first iteration of the outer loop.
Since $\hat{\mathit{pre}}$ tells us that \texttt{i} still has unexplored predecessors, we cannot eliminate, and next explore \texttt{ok} in step~(2).
We then eliminate \texttt{ok}---its only predecessor \texttt{i} is fully explored---but since \texttt{ok} has just a single self-loop, the elimination has no effect.
In step~(3), we have just explored state $4$, which can now be eliminated.
The result is shown as step~(4).
We proceed in the same pattern in steps (5) through~(8).
Then, in step~(9), we fully explore state~$1$.
Now all predecessors of \texttt{i} are fully explored, and we can eliminate both $1$ and \texttt{i}.
For the sake of illustration, let us pick the more complicated ordering and eliminate \texttt{i} first.
The result is shown as step~(10).
Since \texttt{i} is the initial state, we keep it, but redirect all incoming transitions.
We also merge its rewards, which is why $1$ now has a non-zero reward.
Note that we show rationals in \Cref{fig:ExampleAlg}, but our implementation uses floating-point numbers.
Remember that, without $\hat{\mathit{pre}}$, we might have eliminated \texttt{i} too early; after any subsequent exploration of a state in $\set{1, \dots, n}$, we would then have to re-eliminate \texttt{i}.
We finally eliminate $1$ in step~(11) and explore state $\bot$ in step~(12).
At this point, the outer loop terminates; we read $\mathbb{P}(\diamond\: \set{\texttt{ok}}) = \frac{4375}{4376} \approx 0.999771$ and $\mathbb{E}(\blackdiamond\: \set{\texttt{ok}, \bot}) = 1 + \frac{119918}{119793} \approx 2.001043$.
Observe that, at any time, we kept at most 4 explicit states in memory.
We can arbitrarily increase the size of this model by increasing $n$, but will only ever need at most 4 states in memory.

\end{example}

\paragraph{Long-run average rewards.}
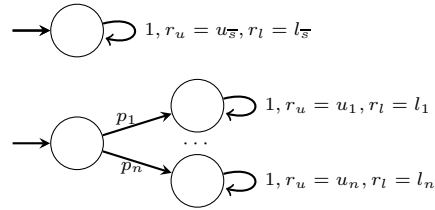
\begin{wrapfigure}[8]{r}{0.5\textwidth}
  \vspace{-5mm}
  \begin{center}
    \begin{tikzpicture}
      \tikzstyle{nodestyle} = [draw, shape = circle, 	inner sep = 0pt, minimum size = 0.7cm];
      \tikzstyle{arrow} = [-stealth, thick];
      \tikzstyle{abovelabel} = [pos = 0.5, above];
      \tikzstyle{belowlabel} = [pos = 0.35, below];
      \def \spacing {0.78cm}
      \def \bendangle {50}
      \def \labelshift {(0.6, 0.1)}
      
      \def \labelshift {(0.0, 0.0)}
      \node (1) [nodestyle] {};
      \node (foo) [left = 0.5cm of 1] {};
      \draw [arrow] (foo)--(1);
      \node (2) [nodestyle, shift={(0, -1.5)}] {};
      \node (bar) [left = 0.5cm of 2] {};
      \draw [arrow] (bar)--(2);
      \node (3) [nodestyle, shift={(1.6, -1)}] {};
      \node (4) [nodestyle, shift={(1.6, -2)}] {};		
      \node (5) [shift={(1.6, -1.5)}] {$\cdots$};
      \path [-stealth, thick]
      (2) edge [] node[left,near end] {$p_1$\ \ } (3);
      \path [-stealth, thick]
      (2) edge [] node[left,near end] {$p_n$\ } (4);
      \path[-stealth, thick]		
      (1) edge [loop right] node {$1, \rfunc_u = u_\init, \rfunc_l = l_\init$} (1);
      \path[-stealth, thick]		
      (3) edge [loop right] node {$1, \rfunc_u = u_1, \rfunc_l = l_1$} (3);
      \path[-stealth, thick]		
      (4) edge [loop right] node {$1, \rfunc_u = u_n, \rfunc_l = l_n$} (4);
      
      \draw [arrow] (foo)--(1);
    \end{tikzpicture}
    \caption{Computation of long-run averages.\label{fig:lra}}
  \end{center}
  \vspace{-10pt}
\end{wrapfigure}
The algorithm we presented so far computed reachability probabilities and expected rewards.
For long-run average reward properties, there are no goal states.
In such a case, our state elimination procedure computes the \emph{recurrence reward} for each BSCC~\cite{GHS18b}.
To obtain the long-run average reward, we need to divide the recurrence reward for the rewards as given in the DTMC by the recurrence reward that we would obtain if all states had reward~$1$.
We can do so by straightforwardly extending \Cref{alg:ExploreEliminate,alg:Eliminate} to work on two reward structures $r_u$ and $r_l$ in parallel.
Upon termination of the outer loop in \texttt{ExploreEliminate}, we then have one of the two situations described at the end of Sect.~4 in~\cite{GHS18b}, and can again directly read off the value for our ($\mathbb{L}$-)property.
Consider \Cref{fig:lra}:
In the simpler case, the remaining model consists of the initial state $\init$ with a self-loop with probability one and $\rfunc_u = u_\init$, $\rfunc_l = l_\init$.
In this case, the average value is $\frac{u_\init}{l_\init}$.
In the other case, the remaining model consists of the initial state $\init$ which has a probability of $p_i$ to move to one of the other $n$ remaining states $s_i$ , $i = 1, \ldots, n$, which all have a self-loop with probability one and $\rfunc_u(s_i) = u_i$, $\rfunc_l(s_i) = l_i$.
In this case, the average value is $ \sum_{i=1,\ldots,n} p_i \frac{u_i}{l_i}$.

When computing long-run average rewards, the final value for $\mathbb{L}$ may be small, but the two recurrence rewards that we need to divide are often extremely large numbers beyond what can usefully be represented as 64-bit (\ie double-precision) floating point numbers.
We thus implemented a variant of our algorithm that uses the GNU MPFR library (see \href{https://www.mpfr.org/}{mpfr.org}) for arbitrary-precision floating-point arithmetic, allowing us to use more than 64 bits.
We did not find this to significantly affect the performance of the overall approach.

\paragraph{Alternatives and optimisations.}
So far, we have assumed that we compute successors for each state explicitly and individually.
For the state elimination phase, doing so is indeed necessary.
However, for just exploring the states we could also compute the \emph{transition relation} as a BDD, and then use the transition relation to symbolically explore the set of reachable states.
This is the standard approach in model checkers such as PRISM~\cite{KNP11} and potentially faster than the semi-symbolic approach we have discussed.
Also, using the transition relation and according MTBDD operations (in particular sum-abstraction), the number of predecessors of each state can also be computed symbolically.

We have so far assumed that the reachable states and the number of predecessors are stored as (MT)BDDs.
An alternative to this approach is to store these numbers on secondary storage (e.g. hard disk) in a similar way as e.g. in~\cite{HartmannsH15}.
This approach would be useful for models the state space of which is not suitable to be stored as a BDD.
This might be the case because of lack of implicit symmetries or because the size of the representation of each state is not constant.

\section{Experimental Evaluation}
\label{sec:Experiments}
We have implemented a preliminary version of our method which we integrated as a plugin for the probabilistic model checker ePMC~\cite{HahnLSTZ14}.
For the analysis, we transform the model and property into C++ code so as to have a means to quickly compute successors of states, similar to the approach used in SPIN~\cite{Holzmann97}.
This C++ file is then appended with code so as to achieve the following:
In the first phase, we then explore the state space in a breadth-first manner where we explore each state explicitly but store sets of states as BDDs, using the BDD package CUDD~\cite{BaharFGHMPS97}.
In the second phase, we generate an MTBDD mapping all states to value $0$.
Then, we iterate over all reachable states, recompute their successors, and increment the value of these successors in the MTBDD by $1$ each time.
In the third phase, we execute the state elimination algorithm as discussed.
This C++ code is then compiled and run in a process separate from ePMC, such that we can measure the memory usage exactly
(the memory usage of ePMC itself is not of much interest, because it is rather small and about the same for any analysis).

In the following, we apply our tool on several case studies from the website of the probabilistic model checker PRISM.
All experiments were performed on a MacBook Pro with a 2.7 GHz Quad-Core Intel Core i7 processor and 16 GB 2133 MHz LPDDR3 RAM.
In the following tables, ``model states'' is the total number of states the model has for the given parameters, ``result'', is the value of the property computed, ``time'' is the total time of the analysis in seconds, ``exp. states'' (``exp. transitions'') are maximal number of states (transitions) being stored explicitly at the same time.
By ``peak mem'' we denote the maximal memory usage in MB of the analysis process.

\subsection{Simple Molecular Reactions: $\mathrm{Na} + \mathrm{Cl} \leftrightarrow \mathrm{Na}^+ + \mathrm{Cl}^-$}
This case study~\footnote{https://www.prismmodelchecker.org/casestudies/molecules.php} is a CTMC modelling the chemical reaction $\mathrm{Na} + \mathrm{Cl} \leftrightarrow \mathrm{Na}^+ + \mathrm{Cl}^-$.
The parameters of this case study are $\mathit{N1}$, the initial number of $\mathit{Na}$ molecules and $\mathit{N2}$, the initial number of $\mathit{Cl}$ molecules.
In Table~\ref{tab:perf-chem}, we consider the performance figures for the analysis of \verb+R=?[S]+ which describes the expected long-run average number of $\mathit{Na}$ molecules.
Here, we consider a starting configuration in which initially the number of $\mathit{Na}$ and $\mathit{Cl}$ is the same, that is, $\mathit{N1} = \mathit{N2}$.

\begin{table}
  \begin{tabular}{rrrrrrrr}
    \toprule
    $\mathit{N1} {=} \mathit{N2}$ & model states & result & time & exp. states & exp. trans & peak mem\\
    \cmidrule{1-1} \cmidrule{2-7}
               10 &            11 & 2.2623e+01 &      8 & 5 & 5 & 22\\
              100 &           101 & 2.3894e+01 &      8 & 5 & 5 & 22\\
            1,000 &         1,000 & 2.4012e+01 &      7 & 5 & 5 & 22\\
           10,000 &        10,001 & 2.4024e+01 &      7 & 5 & 5 & 25\\
          100,000 &       100,001 & 2.4025e+01 &     10 & 5 & 5 & 28\\
        1,000,000 &     1,000,001 & 2.4025e+01 &     32 & 5 & 5 & 27\\
       10,000,000 &    10,000,001 & 2.4025e+01 &    258 & 5 & 5 & 31\\
      100,000,000 &   100,000,001 & 2.4025e+01 &  2,609 & 5 & 5 & 29\\
    1,000,000,000 & 1,000,000,001 & 2.4025e+01 & 18,807 & 5 & 5 & 25\\
    \bottomrule
  \end{tabular}
  \caption{$\mathrm{Na} + \mathrm{Cl} \leftrightarrow \mathrm{Na}^+ + \mathrm{Cl}^-$ performance figures.
    \label{tab:perf-chem}}
\end{table}

As we see, the model scales well for large numbers of molecules and accordingly large state spaces.
The memory usage grows only slowly with increasing model parameters, and the number of states and transitions required to be stored explicitly is constant.

\subsection{Bounded Retransmission Protocol}
The Bounded Retransmission Protocol~\cite{HSV94}\footnote{https://www.prismmodelchecker.org/casestudies/brp.php} is a file transmission protocol.
Files are divided into $N$ packages, each of which is transferred individually.
Data and confirmation packages are sent over unreliable channels, such that they might get lost.
Packages can only be resent a number of $\mathit{MAX}$ times.
In Table~\ref{tab:perf-brp}, we provide performance figures for the analysis of the property \verb+P=?[ F s=5 ]+, that is the probability that the sender does not eventually report a successful transmission.
``$N$'' and ``$\mathit{MAX}$'' are as discussed above, the other numbers are as in the previous case study.

\begin{table}
\begin{tabular}{rrrrrrrr}
  \toprule
  $N$ & $\mathit{MAX}$ & model states & result & time & exp. states & exp. trans & peak mem\\
  \cmidrule{1-2} \cmidrule{3-8}
      64 &    5 &       4,936 & 4.48e-08   &    9 &    12 &    29 & 25\\
      64 &   10 &       9,101 & 1.05e-15   &    9 &    20 &    53 & 25\\
      64 &  100 &      84,071 & 5.03e-153  &    9 &   140 &   517 & 25\\
      64 & 1000 &     833,771 & 3.14e-1526 &   25 &   258 &   986 & 33\\
     128 &   10 &      18,189 & 2.11e-15   &    9 &    20 &    53 & 25\\
     128 &  100 &     168,039 & 1.01e-152  &   10 &   140 &   517 & 32\\
     128 & 1000 &   1,666,539 & 6.28e-1526 &   38 &   514 & 1,978 & 36\\
     256 &   10 &      36,365 & 4.21e-15   &    9 &    20 &    53 & 27\\
     256 &  100 &     335,975 & 2.01e-152  &   15 &   150 &   517 & 28\\
     256 & 1000 &   3,332,075 & 1.26e-1525 &   72 & 1,026 & 3,962 & 39\\
     512 &   10 &      72,717 & 8.42e-15   &    9 &    20 &    53 & 29\\
     512 &  100 &     671,847 & 4.03e-152  &   18 &   140 &   517 & 32\\
     512 & 1000 &   6,663,147 & 2.51e-1525 &  140 & 1,340 & 5,167 & 42\\
    1024 &   10 &     145,421 & 1.69e-14   &   10 &    20 &    53 & 29\\
    1024 &  100 &   1,262,631 & 8.06e-152  &   29 &   140 &   517 & 30\\
    1024 & 1000 &  13,325,291 & 5.02e-1525 &  280 & 1,340 & 5,167 & 43\\
    2048 &   10 &     290,829 & 3.37e-14   &   12 &    20 &    53 & 27\\
    2048 &  100 &   2,687,079 & 1.61e-151  &   50 &   140 &   517 & 30\\
    2048 & 1000 &  26,649,579 & 1.00e-1524 &  552 & 1,340 & 5,167 & 41\\
    4096 &   10 &     581,645 & 6.74e-14   &   17 &    20 &    53 & 27\\
    4096 &  100 &   5,374,055 & 3.22e-151  &   95 &   140 &   517 & 28\\
    4096 & 1000 &  53,298,155 & 2.01e-1524 & 1,151& 1,340 & 5,004 & 39\\
    8192 &   10 &   1,163,277 & 1.35e-13   &   26 &    20 &    53 & 28\\
    8192 &  100 &  10,748,007 & 6.45e-151  &  187 &   140 &   517 & 27\\
    8192 & 1000 & 106,595,307 & 4.02e-1524 &2,385 & 1,340 & 5,004 & 40\\
   16384 &   10 &   2,326,541 & 2.67e-13   &   48 &    20 &    53 & 27\\
   16384 &  100 &  21,495,911 & 1.29e-150  &  392 &   140 &   517 & 28\\
   16384 & 1000 & 213,189,611 & 8.03e-1524 &4,534 & 1,340 & 5,004 & 40\\
   32768 &   10 &   4,653,069 & 5.39e-13   &   91 &    20 &    53 & 26\\
   32768 &  100 &  42,991,719 & 2.58e-150  &  781 &   140 &   517 & 27\\
   32768 & 1000 & 426,378,219 & 1.61e-1523 &9,039 & 1,340 & 5,004 & 41\\
   65536 &   10 &   9,306,125 & 1.08e-12   &   20 &    54 &   156 & 25\\
   65536 &  100 &  85,983,335 & 5.16e-150  &  140 &   503 & 1,362 & 29\\
  131072 &   10 &  18,612,237 & 2.1572e-12 &  312 &    20 &    54 &26\\
  131072 &  100 & 171,966,567 & 1.03e-149  &2,847 &  140 &    503 & 29\\
   
  \bottomrule
\end{tabular}
\caption{Bounded Retransmission Protocol performance figures.\label{tab:perf-brp}}
\end{table}

Compared to the instances of the PRISM website, we have analysed instances with higher parameter numbers $N$ and $\mathit{MAX}$ because our focus was in the scalability of our method.
For comparison, for the first table entry we used the same parameters as the last table entry on the PRISM website.

As we see, we are able to handle instances with several million states with a low memory usage.
Even for higher parameter values for which the number of total states the model consists of is in the millions, we never use more than a few thousand explicit states and transitions and less than 100MB.

\subsection{Wireless Communication Cell}
This case study is a performance model of wireless communication cells~\cite{HMPT00}\footnote{https://www.prismmodelchecker.org/casestudies/cell.php}.
The parameter $N$ describes the number of channels in a cell.
We analyse the property \verb+R{"calls"}=? [ S ]+ which describes the average number of calls in the cell on the long run.
We provide performance figures in Table~\ref{tab:perf-cell}.

\begin{table}
\begin{tabular}{rrrrrrrr}
  \toprule
  $N$ & model states & result & time & exp. states & exp. trans & peak mem\\
  \cmidrule{1-1} \cmidrule{2-7}
         10,000 &       10,001 & 7.00e+01 &     7 & 5 & 5 & 24\\
        100,000 &      100,001 & 7.00e+01 &     9 & 5 & 5 & 24\\
      1,000,000 &    1,000,001 & 7.00e+01 &    24 & 5 & 5 & 25\\
     10,000,000 &   10,000,001 & 7.00e+01 &   183 & 5 & 5 & 26\\
    100,000,000 &  100,000,001 & 7.00e+01 & 1,731 & 5 & 5 & 25\\
  \bottomrule
\end{tabular}
\caption{Wireless Communication Cell performance figures.\label{tab:perf-cell}}
\end{table}

Also for this case study, the approach works fine in that the number of states and transitions to be stored is small as is the peak memory usage.

\subsection{Crowds Protocol}
The Crowds protocol~\cite{RR98}\footnote{https://www.prismmodelchecker.org/casestudies/crowds.php} is a means to allow anonymous web browsing.
To do so, messages are not directly sent, but forwarded to other users, who might either forward them again or send them to the destination.
By doing so, it is hard for attackers to decide whether the sender of a message is the original sender or is just forwarding the message.
The model version we consider has two parameters: $\mathit{TotalRuns}$ is the number of routing paths of the model instance, and $\mathit{CrowdSize}$ is the number of honest participants of the protocol.
We consider the property \verb+Pmax=?[true U (new & runCount=0 & observe0 > observe1 & ... &+\\\verb+observe0 > observe19)]+ which means that an attacker is eventually able to observe the true sender of a message more often than participants just forwarding the message and is thus able to guess the original sender.
In Table~\ref{tab:perf-crowds}, we provide performance figures.

\begin{table}
  \begin{tabular}{rrrrrrrr}
    \toprule
    $\mathit{CrowdSize}$ & $\mathit{TotalRuns}$ & model states & result & time & exp. states & exp. trans & peak mem\\
    \cmidrule{1-2} \cmidrule{3-8}
      5 & 5 &   8,653 & 2.7884e-01 &   9 &    289 &  1,278 & 77\\
      5 & 6 &  18,817 & 2.9791e-01 &   9 &    510 &  2,236 & 77\\
      5 & 7 &  37,291 & 3.1812e-01 &   9 &    823 &  3,898 & 164\\
     10 & 5 & 111,294 & 2.1662e-01 &  19 &  6,604 & 33,513 & 2487\\
     10 & 6 & 352,535 & 2.3162e-01 & 106 & 17,824 & 89,120 & 10698\\
     15 & 5 & 592,060 & 1.9674e-01 & 676 & 44,104 & 356,143 & 9240\\
    \cmidrule{1-2} \cmidrule{3-8}
  \end{tabular}
  \caption{Crowds Protocol performance figures.\label{tab:perf-crowds}}
\end{table}

As we see, for this model the current implementation does not perform very well.
The reason is that too many states and transitions have to be stored explicitly at the same time, leading to a large memory overhead.

\subsection{Embedded Control System}
This case study~\cite{MCT94}\footnote{https://www.prismmodelchecker.org/casestudies/embedded.php} is an embedded control system which features a cyclic polling process.
If a certain component detects that more than a given number $\mathit{MAX\_COUNT}$ of cycles have been skipped due to issues,then the system is shut down for safety reasons.
We consider the property \verb+R{"danger"}=? [ F "down" ]+, which expresses the expected time the system is in an endangered state before it eventually has to be shut down.
We provide performance figures in Table~\ref{tab:perf-ecs}.
Again, the analysis method works fine, as the number of states and transitions stored explicitly is limited.

\begin{table}
\begin{tabular}{rrrrrrrr}
  \toprule
  $\mathit{MAX\_COUNT}$ & model states & result & time & exp. states & exp. trans & peak mem\\
  \cmidrule{1-1} \cmidrule{2-7}
    512 &     320,316 & 3.3454e-01 &   56  & 267 & 11,938 & 80\\
   1024 &     639,804 & 3.3731e-01 &   107 & 267 & 11,938 & 80\\
   2048 &   1,278,780 & 3.4283e-01 &   201 & 267 & 11,938 & 80\\
   4096 &   2,556,732 & 3.5371e-01 &   400 & 267 & 11,938 & 80\\
   8192 &   5,112,636 & 3.7485e-01 &   794 & 267 & 11,938 & 80\\
   16384 & 10,224,444 & 4.1469e-01 & 1,579 & 267 & 11,938 & 80\\
   32768 & 20,448,060 & 7.6563e-01 & 2,914 & 284 & 10,513 & 72\\
  \bottomrule
\end{tabular}
\caption{Embedded Control System performance figures.\label{tab:perf-ecs}}
\end{table}

\section{Conclusion and Future Work}
\label{sec:Conclusion}
In this paper, we have discussed a new memory-efficient analysis method for properties of stochastic models.
Our method is widely applicable to a large variety of stochastic models and properties, though, for conciseness of presentation we have concentrated on DTMCs (with some of the case studies being CTMCs).
Experimental evidence has shown that our approach has the potential to analyse models with millions of states with just a few megabytes.
One advantage of our method is that we directly obtain a precise result:
Our current implementation is based on (variable-precision) floating-point numbers, and computation precision is limited by the properties of this representation.
We could however as well use exact (rational) arithmetic and would obtain exact values without any change in the core algorithms, since all intermediate and final values are rational, though at the cost of increased computation time and memory usage.
Alternatively, we could use interval arithmetic so as to obtain precise upper and lower bounds of the values computed, again, without any change in the algorithm, and with only moderate overhead.
This is in contrast to methods based on value iteration or state-space abstraction, for which special precaution is required to ensure this.
Our current implementation explores and eliminates states in a strict breadth-first search order.
Motivated by the problematic performance of the Crowds protocol case study, we also want to consider different search orders so as to improve the behaviour for models for which the strict breadth-first search order does not perform well.

\newpage



\bibliography{paper}
\bibliographystyle{splncs03}

\end{document}